\documentstyle[12pt,epsf]{article}
\newcommand{\vph}{$\vphantom{\displaystyle\frac{1}{1}}$}
\def\J{$J/\psi$}
\def\j{J/\psi}
\def\X{$\chi$}

\def\P{$\psi'$}
\def\p{\psi'}
\def\U{$\Upsilon$}
\def\u{\Upsilon}
\def\C{c{\bar c}}

\def\b{b{\bar b}}

\def\Q{Q{\bar Q}}
\def\e{\epsilon}

\def\be{\begin{equation}}
\def\ee{\end{equation}}

\def\lsim{\raise0.3ex\hbox{$<$\kern-0.75em\raise-1.1ex\hbox{$\sim$}}}
\def\gsim{\raise0.3ex\hbox{$>$\kern-0.75em\raise-1.1ex\hbox{$\sim$}}}

\def\PL{{ Phys.\ Lett.\ }}
\def\PR{{ Phys.\ Rev.\ }}

\def\PRL{{ Phys.\ Rev.\ Lett.\ }}

\def\ZP{{ Z.\ Phys.\ }}

\begin{document}

\noindent June 1, 2001~ \hfill BI-TP 2001/09

\vskip 1.5 cm

\centerline{\large{\bf Quarkonium Feed-Down and Sequential Suppression}}

\vskip 1.0cm

\centerline{\bf S.\ Digal, P.\ Petreczky and H.\ Satz}

\bigskip

\centerline{Fakult\"at f\"ur Physik, Universit\"at Bielefeld}
\par
\centerline{D-33501 Bielefeld, Germany}

\vskip 1.0cm

\noindent

\centerline{\bf Abstract:}

\medskip

About 40-50 \% of the quarkonium ground states \J(1S) and \U(1S)
produced in hadronic collisions originate from the decay of higher
excitations. In a hot medium, these higher states are dissociated at
lower temperatures than the more tightly bound ground states, leading
to a sequential suppression pattern. Using new finite temperature
lattice results, we specify the in-medium potential between heavy quarks
and determine the dissociation points of different quarkonium states.
On the basis of recent CDF data on bottomonium production, we then obtain
first predictions for sequential \U~suppression in nuclear collisions.

\vskip 1cm

\noindent{\bf 1.\ Introduction}

\bigskip

The large values of the charm and bottom quark masses permit potential
theory to provide a realistic account of quarkonium spectroscopy
\cite{Cornell}-\cite{Eichten}. From these studies, it is known that the
intrinsic length scales of quarkonia are much smaller than those of the
usual hadrons, with $r_{\j} \sim 0.2$ fm and $r_{\u} \sim 0.1$ fm for
the radii of the lowest $\C$ and $\b$ vector mesons, respectively, in
contrast to about 1 fm for light $q \bar q$ state radii. This
``short-distance" nature of quarkonium states suggests that at least
some features of their production in hadronic collisions should be
accessible to perturbative QCD calculations, and that indeed turns out
to be the case.

\par

The simplest and most general model for quarkonium production, the
color evaporation model \cite{CE}, postulates that the cross-section
for the production of a given charmonium or bottonium state is simply a
fixed (energy-independent) fraction of the corresponding perturbatively
calculated $\C$ or $\b$ production cross-section. The resulting
predictions for the energy variation of the \J~and \U~hadroproduction
cross-sections are experimentally very well confirmed, for the \U~over
a range from $\sqrt s \simeq$ 20 to 1800 GeV \cite{Gavai}. The assumed
energy independence of the production ratios of $2S/1S$ and
$3S/1S$ states is found to hold over the same range \cite{Gavai}.

\par

Higher quarkonium excitations decay into lower states with generally
known branching ratios and widths. As an example, the $\psi'=\psi(2S)$
decays into $\j(1S)$ + anything with a branching ratio of 55 \%, after
a mean life-time of more than $10^3$ fm. As a consequence, \J~or
\U~production in hadronic collisions occurs in part through the
production of higher excited states which subsequently decay into the
quarkonium ground states. It is known experimentally that for
both \J~and \U~about 40 - 50 \% of the hadroproduction rate is due to
such feed-down from higher excitations \cite{Cobb} - \cite{CDF99}.

\par

Quarkonium production through feed-down becomes particularly interesting
when quarkonium states are used to probe the hot and dense medium
created in high energy nuclear collisions. It was predicted that
color deconfinement (quark-gluon plasma formation) would lead to
\J~suppression, since sufficiently hot deconfined media dissolve
any $\C$ binding \cite{Matsui}. However, different quarkonium
excitations will dissolve at different temperatures of the medium
\cite{MTM}, and through a lowering of the open charm (beauty) threshold with
temperature, dissociation by $D \bar D$ or $B \bar B$ decay becomes
possible for higher excited states even below the deconfinement point
\cite{DPS1}. Since the life-time of the excitations is much larger than
that of the medium, feed-down production will result in a
characteristic sequential suppression pattern \cite{K-S,G-S}, with the
fraction of \J~or \U~produced through the decay of higher excitations
becoming suppressed at lower temperatures than the directly produced
ground states.

\par

To fully predict this sequential suppression, two prerequisites are
needed. We have to know what fractions of the quarkonium ground state
production originate from which higher excitations, and we have to know
at what temperature or energy density of the hot medium a given
excitation dissolves. The first can be determined either experimentally
or through a viable model for quarkonium hadroproduction. The second is
a well-defined problem for finite temperature lattice QCD studies. At
present, neither problem is completely solved. However, in the case of
charmonium, the production rates of the higher excitations are
experimentally known, and for bottonium production, recent Fermilab
data \cite{CDF99,CDF95} provide the basis for fairly reliable estimates.
This determines the structure (i.e., the sequence and the different
heights) of the various suppression steps for \J~and \U~production,
but not the actual positions of these steps as function of the
temperature or energy density. To estimate these, we make use of recent
lattice studies calculating the temperature behavior of the heavy quark
potential in full QCD \cite{DPS1,Peikert}. As a result, we obtain
a modified \J~suppression pattern \cite{K-S,G-S}, including the decay
of \P~and $\chi_c$ in confined matter \cite{DPS1}, and then first
quantitative predictions for the sequential \U~suppression to be
studied in forthcoming RHIC and LHC experiments.

\par

The structure of this paper is the following. In section 2, we first
summarize the hadroproduction cross sections for the different
charmonium states and then determine the corresponding bottonium cross
sections from the mentioned new CDF data. As a result, we can
fully describe the origin of the \J~and \U~produced in hadronic
collisions in terms of feed-down from higher excited states.
In section 3, we consider the temperature dependence of the heavy quark
potential obtained in recent lattice QCD studies. Solving the
corresponding Schr\"odinger equation, we then determine in section 4 the
dissociation parameters for the different states in a deconfined medium.
This leads to quantitative estimates of the sequential \J~and
\U~suppression patterns.

\bigskip

\noindent{\bf 2.\ Quarkonium Production and Feed-Down}

\bigskip

\noindent{\bf A.\ Charmonium States}

\medskip

It is well known that \J~production in hadron-hadron collisions
is to a considerable extent due to the production and subsequent
decay of higher excited $\C$ states \cite{Cobb}-\cite{Antoniazzi}. We
shall here summarize the situation following systematic studies using
pion and proton beams at 300 GeV incident energy \cite{Antoniazzi}.
In the first two columns of Table 1, we list the cross-sections
$\sigma^d_i$ obtained for the direct production (excluding feed-down)
of the different charmonium states $\psi(1S)$, $\chi(1P)$~and
\P=$\psi(2S)$~in $\pi^-$-nucleon and $p$-nucleon interactions,
normalized to the overall measured \J~cross-section $\sigma_{\j}$,
which includes all feed-down contributions. Hence $R_i(\pi^-N)\equiv
\sigma^d_i(\pi^-N)/\sigma_{\j}(\pi^-N)$ for the directly produced state
$i$ in $\pi^-N$ interactions, and similarly for $p~N$ collisions.

\par

Making use of the branching ratios $B[\chi_1(1P) \to \psi(1S)] = 0.27
\pm 0.02$, $B[\chi_2(1P) \to \psi(1S)] = 0.14 \pm .01$, and
$B[\psi(2S) \to \psi(1S)] = 0.55 \pm 0.05$, one obtains the fractional
feed-down contributions $f_i$ of the different charmonium states to the
observed \J~production; these are shown in the next two columns of
Table 1. Also listed are the dissociation energies $E^i_{\rm dis}$,
\be
E^i_{\rm dis} \equiv 2M_D - M_i.
\label{2.0}
\ee
measuring how far the mass of state $i$ lies below the zero-temperature
open charm threshold $2M_D$=3.740 GeV.

\medskip

\begin{center}
\begin{tabular}{|c||c|c||c|c||c|}
\hline
      &          &              &              &          &     \\
state & $R_i(\pi^-N)$ & $R_i(p~N)$ & $f_i(\pi^-N)$ [\%] & $f_i(p~N)$
[\%] & $E_{\rm dis}$ [MeV] \\
&  &  & & &  \\
\hline
\hline
&  &  & & &  \\
$J/\psi(1S)$ & 0.57 $\pm$ 0.03 & 0.62 $\pm$ 0.04  & 57 $\pm$ 3
& 62 $\pm$ 4 & 0.642  \\
&  &  & & & \\
\hline
&  &  & & &  \\
$\chi_1(1P)$  & 0.72 $\pm$ 0.18  & 0.60 $\pm$ 0.15  & 20 $\pm$ 5 &
16 $\pm$ 4 & 0.229  \\
&  &  & & &  \\
\hline
&  &  & & &  \\
$\chi_2(1P)$ & 1.04 $\pm$ 0.29 & 0.99 $\pm$ 0.29 & 15 $\pm$ 4 & 14
$\pm$ 4
& 0.183  \\
& &  & & & \\
\hline
& &  & & & \\
$\psi(2S)$ & 0.14 $\pm$ 0.04 & 0.14 $\pm$ 0.04 & ~8 $\pm$ 2 & ~8 $\pm$ 2
&
0.054 \\
& &  & & & \\
\hline
\hline
& &  & & & \\
J/$\psi$ & 1 & 1 & 100 & 100 & \\
& &  & & & \\
\hline
\end{tabular}
\end{center}

\par

Table 1: Cross-sections for direct charmonium production in $\pi^-N$ and
$pN$ collisions, normalized to the overall \J~production cross section
in the corresponding reaction \cite{Antoniazzi}; feed-down fractions and
mass gap to the open charm threshold.

\bigskip

From Table 1 it is seen that some 60 \% of the observed \J~are directly
produced, about 30 \% come from \X~and about 10 \% from \P~decay.
According to the color evaporation model, feed-down fractions as well
as cross section ratios are energy-independent. The results shown in
Table 1 for the ratio of \P~to overall \J~production are in excellent
agreement with a variety of experimental results over the range $\sqrt
s = 18 - 65$ GeV, which give $0.14 \pm 0.034$ as average \cite{Gavai}.
Even Tevatron results at $\sqrt s = 1.8$ TeV for charmonium transverse
momenta $p_T \geq$ 5 GeV lead to $0.19 \pm 0.05$ and are thus within
errors in accord with the quoted average value. Similarly, the ratio
of \X~to \J~production is found to be constant over the range $\sqrt s
=7-65$ GeV \cite{Gavai}. As noted, the color evaporation model
postulates that the different charmonium state cross-sections
$\sigma^d_i$ are constant (i.e., energy-independent) fractions $c_i$ of
the overall `hidden charm' cross-section
\be
\sigma_i^d(s) = c_i \sigma_{\C}(s) ~~{\rm with}~~M_{\C} \leq 2M_D.
\label{2.1}
\ee
This assumption is thus indeed very well satisfied, even though the
corresponding production cross-sections themselves vary
considerably in the energy range in question, in accord with the
perturbatively calculable variation of $\sigma_{\C}(s)$. Note that
$\sum_i c_i << 1$, since more than half of the $\C$ pairs formed with
$M_{\C} < 2M_D$ acquire the missing energy from the color field and then
contribute to open charm production.

\par

The color evaporation model does not predict the values of the factors
$c_i$. More detailed arguments do, however, lead to some relations
between the different cross-sections. Projecting a color singlet $\C$
state onto different quantum number configurations leads to the
estimate \cite{Schuler}
\be
{\sigma^d(2S) \over \sigma^d(1S)} \simeq {\Gamma (\psi(2S)\to e^+e^-)
\over
\Gamma(\psi(1S) \to e^+e^-)} \left( {M_{\j} \over M_{\p}}\right)^3
\simeq 0.24,
\label{2.2}
\ee
where $\Gamma$ denotes the corresponding dilepton decay width of the
state in question, $M$ its mass. The values $ 0.23 \pm 0.06$ and $0.22
\pm 0.05$ obtained from Table 1 for $\pi^-N$ and $pN$ collisions,
respectively, agree well with relation (\ref{2.2}).

\par

The ratios between the different $\chi_l(1P)$ states are predicted to be
governed essentially by the orbital angular momentum degeneracy
\cite{orbital}; we thus expect for the corresponding cross-sections
\be
\chi_0(1P) : \chi_1(1P) : \chi_2(1P) = 1 : 3 : 5.
\label{2.3}
\ee
From Table 1 we have for $\pi^-N$ collisions $\chi_2(1P) / \chi_1(1P)
\simeq 1.44 \pm 0.38$ and thus reasonable agreement with the predicted
ratio 1.67. For $pN$ interactions, the experiment measures only the
combined effect of $\chi_1$ and $\chi_2$ decay (30 \% of the overall
\J~production); the listed values are obtained by distributing this in
the ratio 3:5.

\bigskip

\noindent {\bf B.\ Bottonium States}

\bigskip

Production cross-sections for the different $(nS)$ bottonium below the
open beauty threshold are known over a considerable energy range
\cite{Gavai,CDF95}, and as for charmonia, the resulting ratios are with
very good precision energy-independent \cite{Gavai}. Corresponding
cross-sections for the sub-threshold $(nP)$ states are so far available
only for transverse momenta $p_T\geq 8$ GeV/c \cite{CDF99}. To analyse
the complete feed-down pattern, we thus have to find a way to
extrapolate these data to $p_T=0$.

\par

\begin{figure}[h]
\vspace*{1cm}
\epsfxsize=9cm
\epsfysize=9cm
\centerline{\epsffile{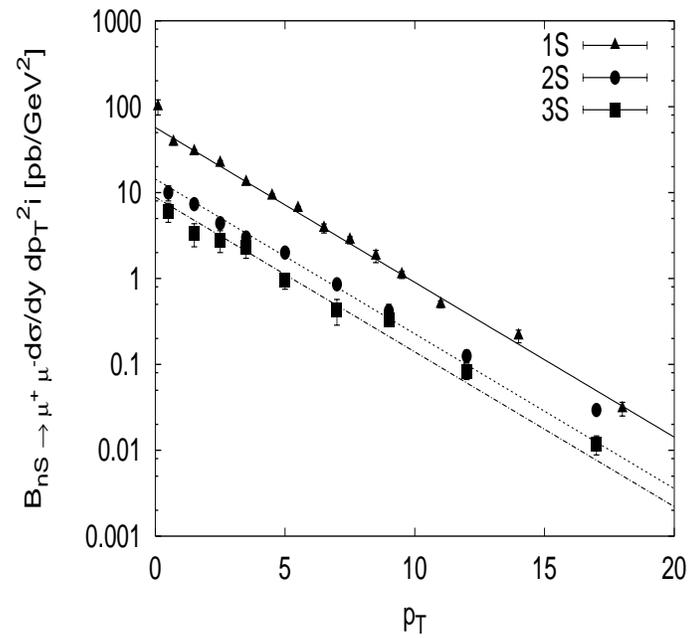}}
\vspace*{-0.3cm}
\caption{The transverse momentum dependence of the inclusive
production cross sections for different ($nS$) bottomonium states; 
the lines are exponential fits.}
\label{dsig}
\end{figure}

In Fig.\ \ref{dsig}, we show the $p_T$ distributions of the \U, \U' and
\U'' states. It is evident that they exhibit a very similar transverse
momentum behavior, which can be parametrized quite well by the form
\be
{d \sigma^i \over dy dp_T^2} = N_i \exp \{(-0.415)p_T\},
\label{2.4}
\ee
where $N_i$ specifies the overall cross-section values of the different
states $i$. We shall therefore now assume that all bottonium states,
$nP$ as well as $nS$, are governed by the same $p_T$ distribution, so
if we know a production ratio in one $p_T$ inteval, we know the ratio
of the overall cross-sections. To test the validity of assuming such a
universal $p_T$ dependence, we have compared data for the different $nS$
states taken in the interval $1 \leq p_T \leq 18$ GeV to an
extrapolation of corresponding data for $8 \leq p_T \leq 18$ GeV.
The resulting difference of about 10 \% indicates the uncertainty
inherent in our procedure.

\par

To complete the feed-down analysis, we need the decay branching ratios
from higher to lower bottonium states, which can be readily computed
from the Particle Data Compilation. From the measured cross-sections
including feed-down effects (denoted by $\sigma$) we now want to
reconstruct the direct production cross-sections (denoted by
$\sigma^d_i$) of all sub-threshold bottonium states $i$. This leads to
the relations
\be
\sigma(\u) = \sum_i B[i \to \u]~ \sigma^d_i,
\label{2.5}
\ee
\be
\sigma(\u') = \sum_i B[i \to \u']~ \sigma^d_i,
\label{2.6}
\ee
and
\be
\sigma(\u'') = \sigma^d(3S)
\label{2.7}
\ee
for the three different $\b$ $S$ states. The branching ratios $B[i \to
\u]$ and $B[i \to \u']$ are compiled in \cite{Braaten}.

\par

The experiment \cite{CDF99} further provides the fractions $F_{\u}^{1P}$
and $F_{\u}^{2P}$ of \U~production coming from the different $1P$ and
$2P$ $\chi_b$ states. For these fractions we know
\be
F^{1P}_{\u} \sigma(\u) = \sum_i B[i \to (1P)]B[(1P) \to \u]~ \sigma_i^d
\label{2.8}
\ee
and
\be
F^{2P}_{\u} \sigma(\u) = \sum_i B[i \to (2P)] B[(2P) \to \u]~ \sigma_i^d
\label{2.9}
\ee
in terms of the direct production cross sections for the different
states. The branching ratios are again given in \cite{Braaten}.
Eqs.\ (\ref{2.8}) and (\ref{2.9}) illustrate that the observed
$\chi_b$ states which can decay into \U~arise themselves in part through
feed-down from still higher excited states. We have to specify the
different fractions for this, since e.g.\ the melting of the $\u(2S)$
will also remove that fraction of the observed $\chi_b(1S)$ production
which comes from $\u(2S)$ decay.

\par

The experimental values for $F^{1P}_{\u}$ and $F^{2P}_{\u}$ are given in
\cite{CDF99} for transverse momenta $p_T \geq 8$ GeV/c. Since the
different $S$ states show a universal $p_T$ dependence, we assume
the same to hold for the $P$ states and thus take the measured values
\be
F^{1P}_{\u} = 0.27 \pm 0.11~~~,~~~ F^{2P}_{\u} = 0.11 \pm 0.06
\label{2.10}
\ee
to remain applicable for the entire $p_T$ range. Making use of the
measured overall cross sections $\sigma_{\u}=26.9 \pm 0.6$ mb,
$\sigma_{\u'}=13.1 \pm 0.5$ mb and $\sigma_{\u''}=5.5 \pm 0.5$ mb,
we
can then solve the five equations (\ref{2.5}) - (10) for the different
direct production cross sections. The results are shown in Table 2,
normalized to $\sigma_{\u}$, together with the relative fractions each
state contributes to the overall \U~production. For the $P$-states
$\chi_b$, the shown values are based on the overall cross-section
$\sigma(\chi_b) = \sigma(\chi_{b0}) + \sigma(\chi_{b1}) +
\sigma(\chi_{b2})$, with the three orbital states assumed to contribute
in the ratios 1:3:5 \cite{orbital}. Again we also list the values
\be
E_{\rm dis} = 2M_B - M_i
\label{2.11}
\ee
of the corresponding zero-temperature dissociation energies.

\begin{center}
\begin{tabular}{|c||c||c||c||c|}
\hline
      &           &          &    & \\
state & $R_i({\bar p}p)$ & $f_i({\bar p}p)$ [\%] &
$E_{\rm dis}$ [GeV] & ${f_i (\bar p p)}_{\rm NRQCD}$ [\%]\\
& & &  &\\
\hline
\hline
& & & & \\
$\u(1S)$ & 0.52 $\pm$ 0.09 & 52 $\pm$ 9  & 1.098 & $0.52 \pm 34$\\
& & & & \\
\hline
& & & & \\
$\chi_b(1P)$ & ~1.08 $\pm$ 0.36 & 26 $\pm$ 7  & 0.670 & $0.24 \pm 8$\\
& & & & \\
\hline
& & & &\\
$\u(2S)$ & ~0.33 $\pm$ 0.10 & 10 $\pm$ 3 & 0.535 & $8 \pm 7$\\
& & & &\\
\hline
& & & &\\
$\chi_b(2P)$ & ~0.84 $\pm$ 0.4 & 10 $\pm$ 7 & 0.305 & $14 \pm 4$\\
& & & & \\
\hline
& & & &\\
$\u(3S)$ & ~0.20 $\pm$ 0.04 & ~2 $\pm$ 0.5 & 0.203 & $2 \pm 2$ \\
& & & &\\
\hline
\hline
& & & &\\
\U & 1 & 100 & & 100\\
& & & & \\
\hline
\end{tabular}\end{center}

\bigskip

\noindent
Table 2: Cross-sections for direct bottomonium production in $\bar{p}-p$
collisions, normalized to the overall \U~production cross section
\cite{CDF99,CDF95}; feed-down fractions and mass gap to the open bottom
threshold; feed-down fractions obtained in NRQCD.

\bigskip

For the feed-down pattern in \U~production we thus find approximately 50
\% direct $\u(1S)$ production, 30 \% from direct $\chi_b(1P)$, 10 \%
from direct $\u'(2S)$ and 10 \% from direct $\chi_b(2P)$ production and
subsequent decay into $\u(1S)$. Again the energy independence of these
fractions, as required by the color evaporation model, is well
satisfied for the measured ratios \cite{Gavai}. To test the consistency
of the feed-down pattern with more recent theoretical considerations,
we have also calculated the fractions for the whole measured $p_T$
interval using the NRQCD factorization formula \cite{Bodwin,Braaten}.
The results are presented in the last column of Table 2; the details of
the calculations are given in the appendix.

\bigskip

\noindent{\bf 3.\ The Heavy Quark Potential in Hot Media}

\bigskip

In finite temperature lattice QCD, the temperature behavior of the
static $\Q$ potential $V(T,r)$ is obtained from Polyakov loop
correlations measuring the free energy $F(T,r)$,
\be
- T~\ln <L(0)L^+(r)> = F(T,r) + C  =  V(T,r) - TS + C,
\label{3.7}
\ee
where $S$ denotes the entropy due to the introduction of an unbound $\C$
or $\b$ pair into the medium and $C$ the (undetermined) Polyakov loop
normalization. This can in principle be fixed by requiring that at very
short distances, $r << T^{-1}$, the potential has the purely Coulombic
form $\alpha/r$, since in the limit $r \to 0$, the effects of the
medium should become negligible. At present, however, lattice
calculations for high temperatures are probably not yet precise enough
to reach the small $r$ range required; for $T \leq T_c$, the
normalisation is found to be more reliable \cite{DPS1}. We therefore
here leave $C$ open; as will be seen, this does not affect the
determination of the dissociation points of the different bound states;
it only prevents a reliable determination of the binding energy.
The definition of $S$ in Eq.\ (\ref{3.7}) assumes that for $r \to
\infty$, $V(T,r) \to 0 ~\forall~T$. We thus obtain $V(T,r)$ from the
relation
\be
V(T,r) = - T~\ln \left({<L(0)L^+(r)>\over <L^2>}\right),
\label{3.7a}
\ee
where $<L^2>$ denotes the value of $<L(0)L^+(r)>$ for $r \to \infty$.

\par

\begin{figure}
\vspace*{-0.5cm}
\epsfxsize=9cm
\epsfysize=9cm
\centerline{\epsffile{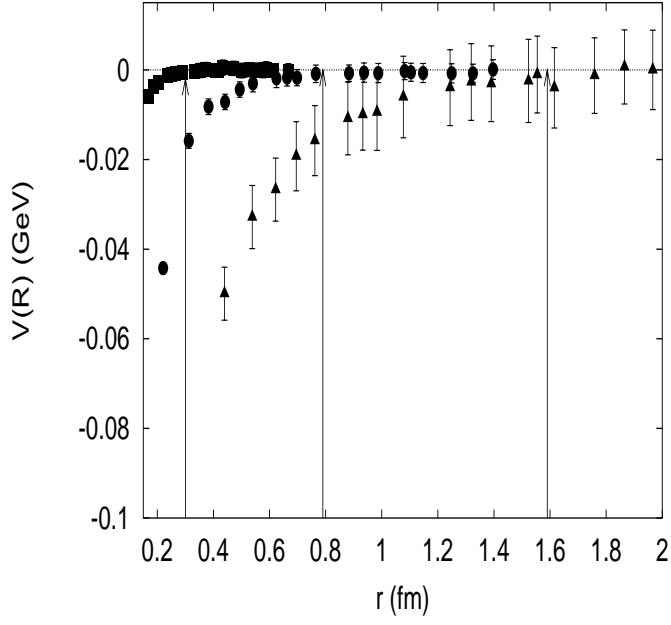}}
\vspace*{-0.3cm}
\caption{The color averaged potential at $T=1.03~ T_c$ (triangles), 
$T=1.45~T_c$ (circles) and $T=3.84~T_c$ (squares). The vertical lines 
indicate the points $r_0$ beyond which the potential becomes 
$r$-independent.}
\label{free}
\end{figure}

\begin{figure}
\epsfxsize=9cm
\epsfysize=9cm
\centerline{\epsffile{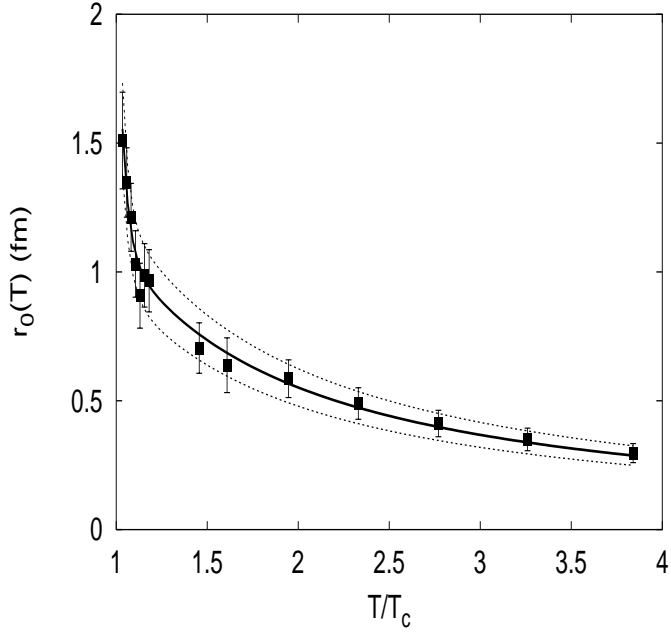}}
\caption{The temperature dependence of the points $r_0$ beyond which
 the potential becomes $r$-independent. The solid line represents the
fit to the data on $r_0(T)$. The dashed lines represent the error band
and were estimated by fitting the data on $r_0$ shifted up (down) by
one standard deviation.}
\label{r_0}
\end{figure}

The free energy (\ref{3.7}) was recently studied on $16^3 \times 4$
lattices for 3 and 2+1 flavor QCD using improved gauge and staggered
fermion actions \cite{Peikert,Laermann}. The quark masses used in these
studies were $m/T=0.4$ for 3 flavor and $m_{u,d}/T=0.4$ and $m_s/T=1$
for 2+1 flavor case. In our analysis we use the 3 flavor potential, for
which the analysis is the most complete. However, we have verified that
differences between the potentials as functions of $r T$ calculated
in 2+1 and 3 flavor cases are in fact small. The resulting potential
in 3 flavor QCD (\ref{3.7a}) for $T > T_c$ is shown in Fig.\ \ref{free}
for some representative temperatures. It is seen that beyond a certain
separation distance $r=r_0(T)$, the free energy of the heavy quark
system becomes a constant, indicating that within the accuracy of the
calculation, the $\Q$ interaction potential vanishes. The resulting
$r_0(T)$ as function of the temperature is given in Fig. \ref{r_0}.
Since for $r > r_0(T)$, there is no more interaction between the static
colour sources, $r_0(T)$ provides a natural limit to all bound state
radii.

\par

\begin{figure}
\epsfxsize=9cm
\epsfysize=9cm
\centerline{\epsffile{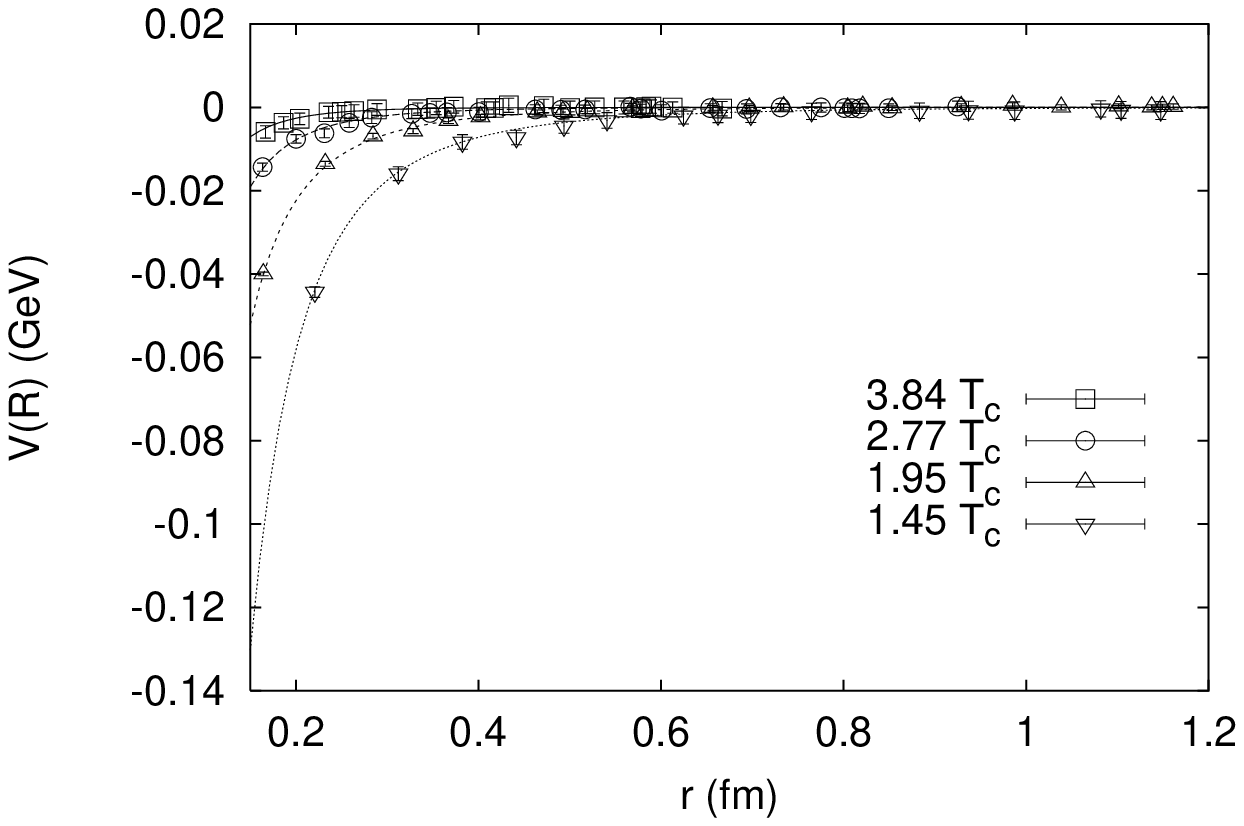}}
\caption{The color averaged potential for temperatures $T \ge 1.45~T_c$;
the lines show the fits described in the text.}
\label{pot-high}
\end{figure}

\begin{figure}
\epsfxsize=9cm
\epsfysize=9cm
\centerline{\epsffile{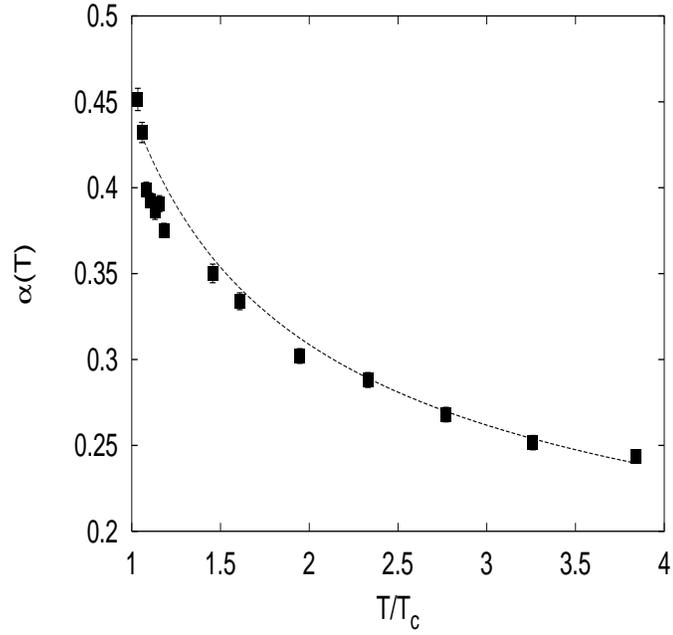}}
\caption{The temperature dependence of the coupling constant;
the line is a fit using the 1-loop running coupling constant formula.}
\label{alpha}
\end{figure}

\begin{figure}
\centerline{
\epsfxsize=9cm \epsfysize=9cm \hskip 0.5cm \epsffile{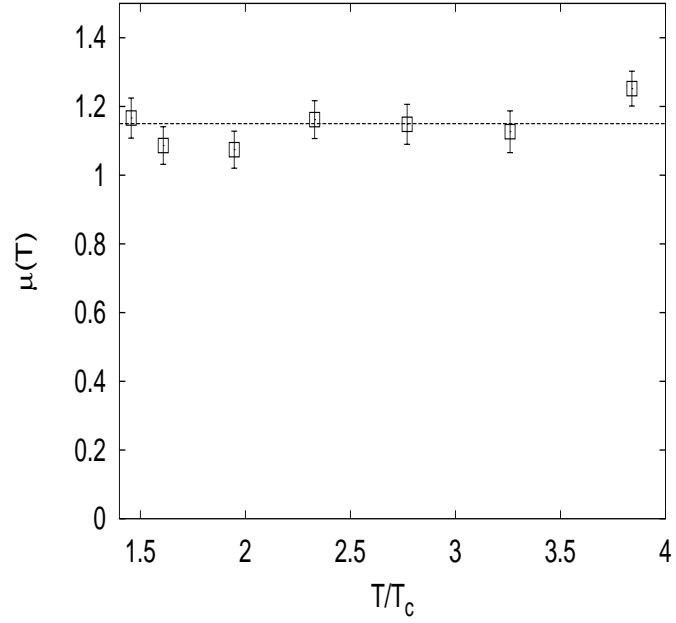} }
\caption{The temperature dependence of the screening mass;
the line shows the average value.}
\label{mu}
\end{figure}

The potential $V(T,r)$ obtained from Eq.\ (\ref{3.7a}) is the average
over color singlet and color octet contributions; it can be written in
the form
\be
V(T,r) = -T ~\ln \left\{ {1\over 9}\exp[-V_1(T,r)/T] +
{8\over 9} \exp[-V_8(T,r)/T] \right\},
\label{3.8}
\ee
where $V_1(T,r)$ and $V_8(T,r)$ specify the singlet and octet
contributions, respectively. In perturbation theory, the leading terms
for both are at high temperature and small $r$ ($r << T^{-1}$) of
Coulombic form,
\be
V_1(T,r) = -{4\over 3} {\alpha(T) \over r}, ~~~
V_8(T,r) = +{1\over 6} {\alpha(T) \over r},
\label{3.9}
\ee
with $\alpha(T)$ for the temperature-dependent running coupling.
In the region just above the deconfinement point $T=T_c$, there will
certainly be significant non-perturbative effects of unknown form.
We therefore first consider the high temperature regime, which we
somewhat arbitrarily define as $T \ge 1.45~T_c$. In this region, we attempt
to parameterize the existing non-perturbative effects through a
conventional screening form, replacing Eq.\ (\ref{3.9}) by
\be
-{3\over 4}V_1(T,r) = 6~V_8(T,r) = {\alpha(T) \over r} \exp\{-\mu(T)r\},
\label{3.10}
\ee
where $\mu(T)$ denotes the effective screening mass in the deconfined
medium.\footnote{The form (\ref{3.10}) is actually only valid for $r
\gg 1/T$. For the distances $r \leq 1/T$ most relevant for quarkonium
studies, the screening is determined by the momentum-dependent
self-energy $\Pi_{00}(\omega=0,p)$ \cite{Gale}. In coordinate space
this leads to an $r$-dependent effective screening mass
\cite{Petreczky}. In our case this $r$-dependence of the effective
screening mass is not important because of the insufficient accuracy of
the lattice results, which in 3 flavor QCD can be fitted well with a
constant effective screening mass.} In Fig.\ \ref{pot-high} we show
the fits obtained with the form (\ref{3.8}/\ref{3.10}), assuming
$\alpha(T)$ and $\mu(T)$ to be unknown functions of $T$. The functional
form of the potential is seen to be reproduced very well, with the
values of $\alpha(T)$ and $\mu(T)$ as given in Figs.\ \ref{alpha} and
\ref{mu}. We conclude that in the high temperature regime, a
perturbative description modified by color screening gives an excellent
account for the lattice results. The screening mass becomes a constant
in units of the temperature,
\be
\mu(T) = (1.15 \pm 0.02)  ~T.
\label{3.11}
\ee
A similar behavior of the screening mass was found in pure SU(2)
and SU(3) gauge theory \cite{heller95} - \cite{cucchieri01}.

\par

We shall now assume that the form (\ref{3.11}) of the screening mass
continues to remain valid as we lower the temperature to $T_c$. Such a
constant screening mass down to $T_c$ is again expected from studies of
pure gauge theory. There the screening mass determined from the color
averaged potential decreases for $T<1.5~T_c$ as $T \to T_c$
\cite{kaczmarek00}; however, the screening mass determined from the
color singlet potential appears to be temperature independent for $T <
10~T_c$ \cite{heller95} - \cite{cucchieri01}. 

\par

\begin{figure}
\epsfxsize=9cm
\epsfysize=9cm
\centerline{
\epsffile{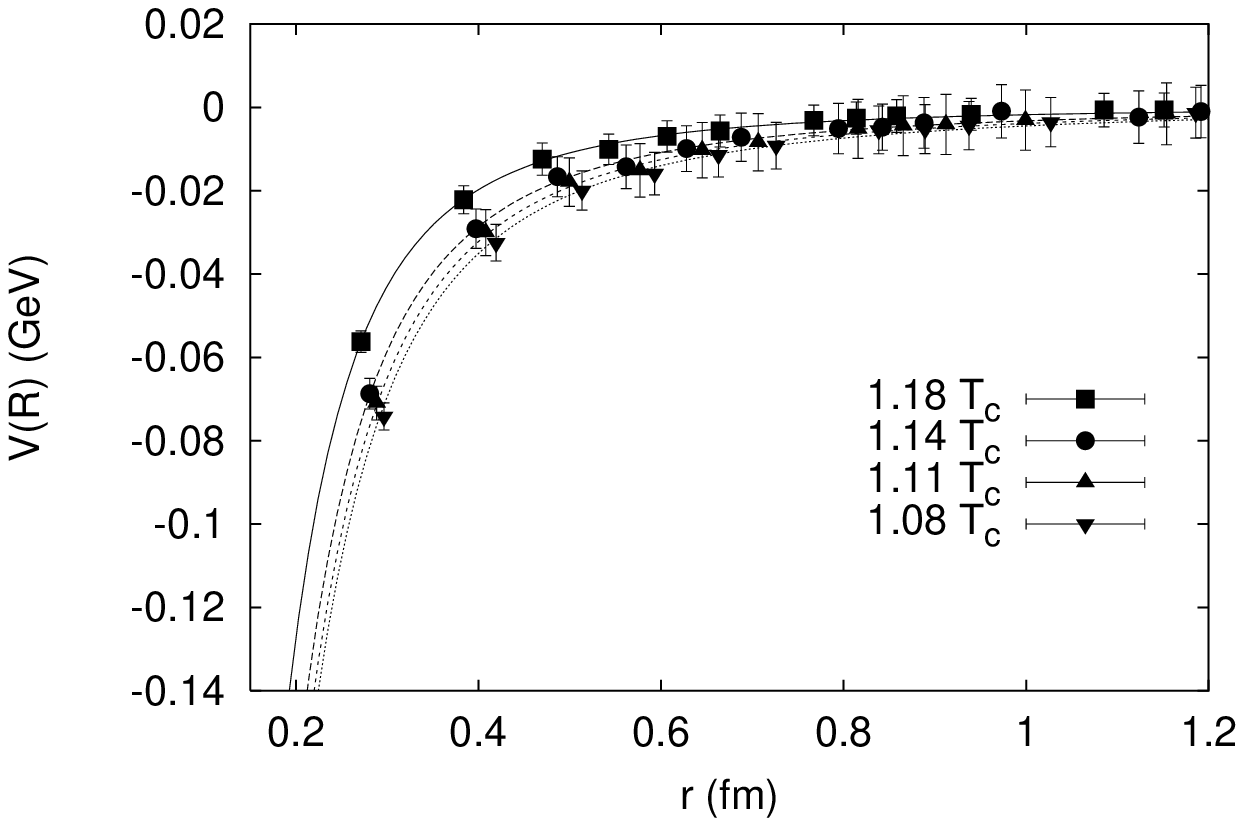} }
\caption{
The color averaged potential for low temperatures $T_c<T<1.45~T_c$;
the lines show the fits described in the text.}
\label{pot-low}
\end{figure}

\begin{figure}
\epsfxsize=9cm
\epsfysize=9cm
\centerline{
\epsffile{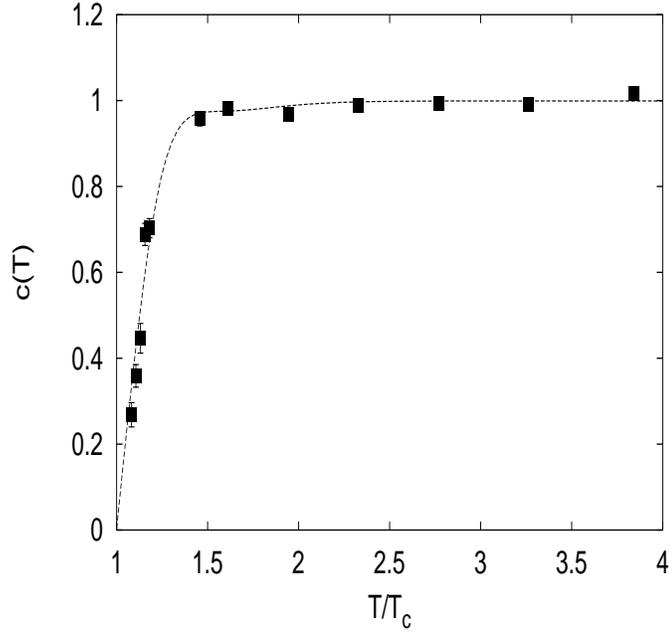} }
\caption{
The ratio $-8 V_8/V_1$ as function of the temperature;
the line is a fit.}
\label{c(T)}
\end{figure}

On the other hand, quenched QCD (pure SU(3) gauge theory) studies
indicate that when $T$ is lowered to $T_c$, the perturbative ratio
$V_1/V_8 = - 8$ will increase in favor of the singlet potential
\cite{attig88}. We therefore try to describe the behavior just above
$T_c$ by a potential of the form (\ref{3.8}), in which the color octet
potential is given by
\be
V_8(T,r) = {c(T)\over 6} {\alpha(T) \over r} \exp\{-\mu r\}
\label{3.12}
\ee
instead of Eq.\ (\ref{3.10}); the factor $c(T) \leq 1$ accounts for the
expected reduction of octet interactions as $T \to T_c$. In the interval
$T_c < T <1.45~T_c$ we thus fit the lattice results for $V(T,r)$ in terms of
the two parameters $c(T)$ and $\alpha(T)$, with $\mu(T)$ given by Eq.\
(\ref{3.11}). In Fig.\ \ref{pot-low} it is seen that this in fact leads
to an excellent parameterization of the lattice results \cite{Peikert}.
The resulting behavior of $c(T)$ is shown in Fig.\ \ref{c(T)}; that of
$\alpha(T)$ is included in Fig.\ \ref{alpha}.  From these
considerations, we obtain the form (\ref{3.10}) for the color singlet
potential, in which $\mu(T)$ is given by Eq.\ (\ref{3.11}) and
$\alpha(T)$ by Fig.\ \ref{alpha}.

\bigskip

\noindent{\bf 4.\ Quarkonium Dissociation in Hot Media}

\bigskip

In the absence of any medium, the masses and radii of the
different charmonium and bottonium states are quite well described by
non-relativistic potential theory \cite{Cornell}-\cite{Eichten}, based
on the Schr\"odinger equation
\be
\left[ 2m_a + {1\over m_a}\nabla^2 + V_1(r) \right] \Phi_i^a = M_i^a
\Phi_i^a,
\label{4.1a}
\ee
where $a=c,b$ specifies charm or bottom quarks, $i$ denotes the
quarkonium state in question, and $r$ is the separation of the two heavy
quarks. Above the deconfinement point, the string tension vanishes and
we are left with the singlet potential $V_1(T,r)$ as determined in the
previous section. We therefore calculate the bound state radii of the
different states, using Eq.\ (\ref{3.10}) in the Schr\"odinger equation.
The radii thus obtained then have to be compared to the limiting
binding radius $r_0(T)$ in order to specify the dissociation points of
the different states.

\par

Before doing this, we have to specify the temperature scale $T_c$. In
the physically interesting case of 2+1 flavor and physical quark masses
it is not very well determined. In the chiral limit of 2 flavor and 3
flavor QCD, one finds $T_c=(173 \pm 16 )$ MeV (2 flavor) and $T_c=(154
\pm 16)$ MeV (3 flavor), where the quoted errors are the sum of
statistical and estimated systematic discretization errors
\cite{Peikert}. For finite u and d quark masses, the critical
temperatures coincide in 2+1 and 2 flavor QCD. We therefore assume that
$T_c=173$ Mev is the relevant critical temperature and use it in our
calculations. We have also studied the results obtained with the
three-flavor temperature $T=154$ Mev; the difference turned out to be
negligable within the present accuracy of our approach.

\par

\begin{figure}
\epsfxsize=9cm
\epsfysize=9cm
\centerline{\epsffile{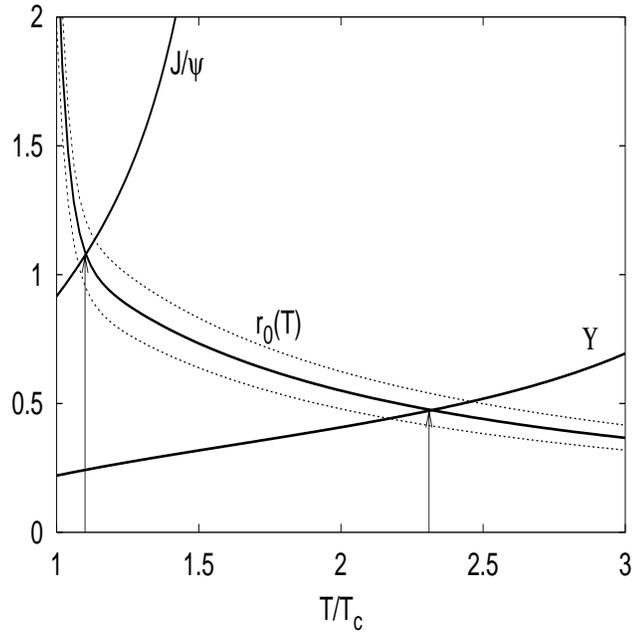}} 
\caption{The radii (in fm) of $J/\psi$ and $\Upsilon$ states as 
function of $T/T_c$, compared to $r_0(T)$. }
\label{radii}
\end{figure}

In Fig.\ \ref{radii}, we illustrate the determination of the
dissociation points by comparing the radii of the charmonium and
bottonium ground states with the limiting binding radius $r_0(T)$. It
is seen that the \J~is dissociated at $T\simeq 1.1~T_c$, the \U~at
$T\simeq 2.5~T_c$. These results are remarkably consistent with those
obtained previously \cite{K-S} using a screened Cornell potential
together with lattice estimates for the screening mass.

\par

We now want to compare the various thresholds for quarkonium
dissociation. In Table 3, we list the different charmonium states,
together with the corresponding dissociation temperatures. For the \J,
this is the value $T\simeq 1.1~T_c$ determined above. The $\chi_c$ and
\P~radii exceed $r_0(T_c)$, so that these two bound states cannot exist
for $T \geq T_c$; the same holds for the $\chi_b(2P)$ and the $\u(3S)$
states. On the other hand, the dissociation points of the $\chi_b(1P)$
and the $\u(2S)$ states coincide approximately with that of the \J; all
three values are here found to be slightly above $T_c$. Bearing in mind
the unknown systematic errors of the method used here, based on the
intersection of $r_0(T)$ and the bound state mass calculated with
a screened Coulombic potential, it seems possible only to conclude that
these states are dissociated very close to the deconfinement point
$T_c$. This is also consistent with the conclusions reached in the
study of the dissociation pattern for $T \to T_c$ from below. The \U,
however, clearly persists up to temperatures well above $T_c$.
In Table 3, we also list the screening masses relative to the
deconfinement value $\mu(T_c)$ for those states which survive beyond
$T_c$ and are then dissociated by color screening; this provides an
indication of the change in the effective screening radius.

\par

Let us next address briefly the survival of those states (\P, $\chi_c$,
\U'', $\chi_b'$) which cannot exist for $T \geq T_c$. In earlier studies
\cite{MTM,K-S}, it had been assumed that they are dissociated at
$T=T_c$. However, a recent analysis \cite{DPS1}, based on the same
lattice study as used here \cite{Peikert}, shows that they will in fact
decay into open charm/beauty states already below $T_c$. Such a decay
becomes possible because the open charm/beauty threshold in a hot
confining medium decreases with temperature faster than the masses of
the corresponding quarkonium states. The relevant decay temperatures
were determined in \cite{DPS1} and are included in Table 3.

\par

\begin{figure}
\epsfxsize=9cm
\epsfysize=11cm
\centerline{\epsffile{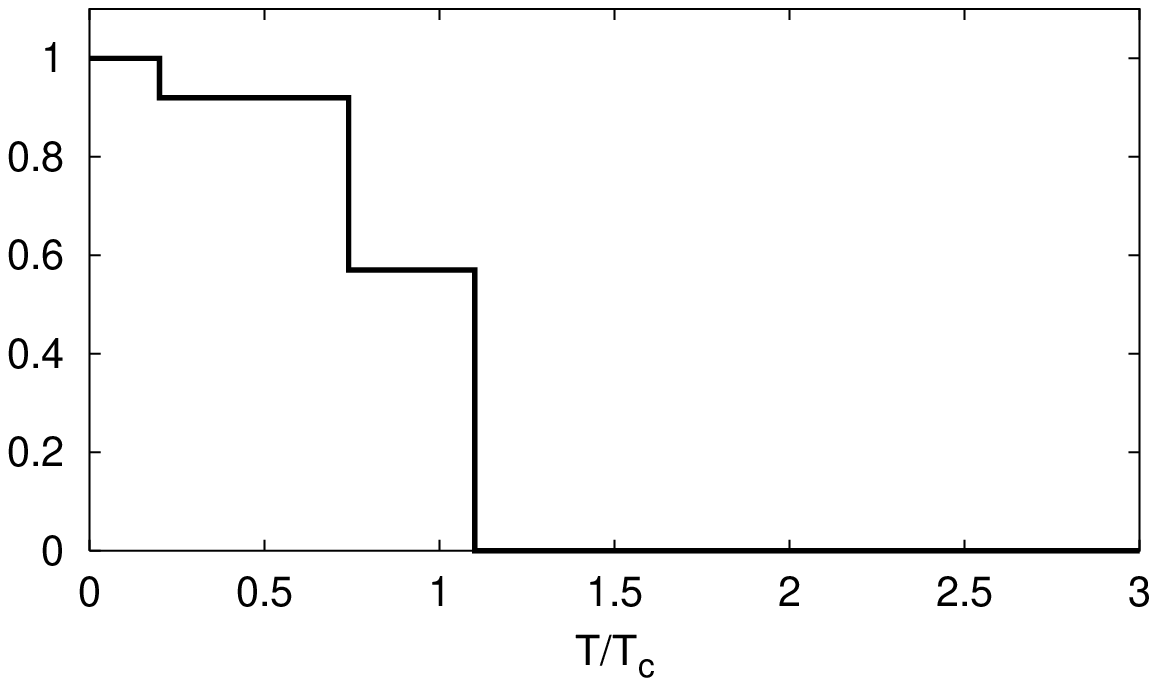}}
\caption{
The \J~suppresssion pattern.
}
\label{psi}
\end{figure}

\begin{figure}
\epsfxsize=9cm
\epsfysize=9cm
\centerline{\epsffile{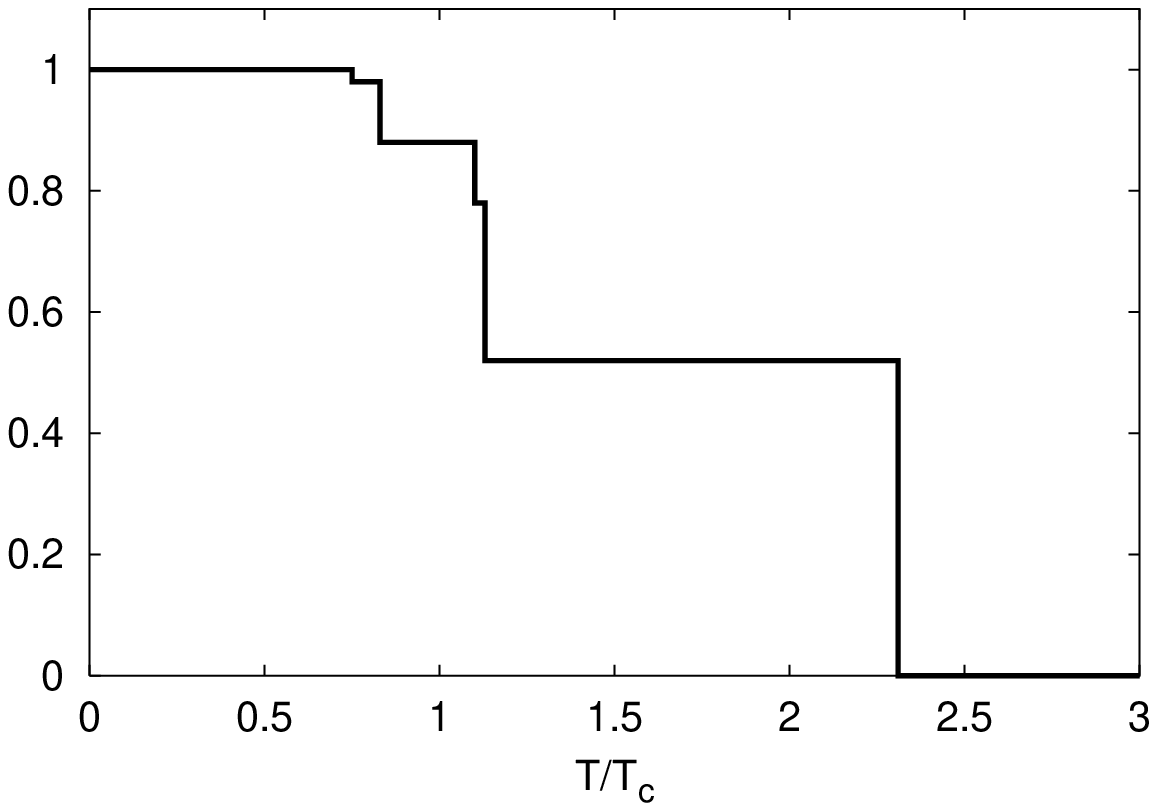}}
\caption{
The \U~suppresssion pattern.
}
\label{upsi}
\end{figure}

Using the results of Table 3 together with the feed-down fractions of
Tables 1 and 2, we obtain the suppression patterns shown in
Figs. \ref{psi} and \ref{upsi}. The new analysis of quarkonia in
confined media has thus led to a modified three-step pattern for
\J~suppression; only the suppression of directly produced \J~requires
the onset of deconfinement. For the \U, we obtain the multi-step form
shown in Fig.\ \ref{upsi}. Here again, two states (\U'', $\chi_b'$)
decay below $T_c$; the next two, \U' and $\chi_b$, are dissociated at or
just above $T_c$, and only the \U~survives much further.

\par

Lattice studies of the kind used here to obtain the heavy quark
potential can in principle also determine directly the energy density
$\e$ of the medium at each temperature, so that we should be able to
give the dissociation points as well in terms of $\e$. At present,
however, various uncertainties (the precision of the $T_c$
determination, the quark mass dependence of $T_c$ and of $\e$,
dependence of $\e$ on the number of flavours, finite lattice effects)
lead to uncertainties of about a factor two in $\e$. A precise
determination of this quantity is thus evidently one of the main
reasons for insisting on increased computer performance in finite
temperature lattice QCD studies; to illustrate: a 10 \% error in $T_c$
leads to a 50 \% error in $\e(T_c)$.

\begin{center}
\baselineskip=1.cm

\begin{center}
\baselineskip=1.cm

\begin{center}
\begin{tabular}{|c||c|c|}
\hline
\vph$q\bar{q}$&$T/T_c$ & $\mu(T_c)/\mu(T)$ \\
\hline
\hline
\vph$J/\Psi$&1.10 & 0.91 \\
\hline
\vph$\chi_c (1P)$&  0.74 & * \\
\hline
\vph$\psi(2S)$&  0.1-0.2 & * \\
\hline
\vph$\Upsilon(1S)$& 2.31  & 0.43 \\
\hline
\vph$\chi_b (1P)$&1.13 & 0.88 \\
\hline
\vph$\Upsilon (2S)$& 1.10 & 0.91 \\
\hline
\vph$\chi_b (2P)$& 0.83& * \\
\hline
\vph$\Upsilon (3S)$& 0.75& * \\
\hline
\end{tabular}\end{center}
\end{center}
\vspace*{0.5cm}

\end{center}

Table 3: The dissociation parameters of different quarkonium states,
as obtained by color screening for $T>T_c$ (present work) and through
decay into open charm/beauty for $T<T_c$ \cite{DPS1}.

\medskip

\medskip

\noindent {\bf Conclusions}

\medskip

In the present work, we have studied quarkonium dissociation by color
screening in a deconfining medium. Using new lattice results
on the color-averaged potential in full QCD, we have determined
the temperature dependence of the color singlet potential. Solving
the Schroedinger equation for heavy $Q \bar Q$ bound state with the
extracted color singlet potential, we have specified the temperatures
at which different bound states dissolve. Combining the results of this
analysis with those recently found for quarkonium dissociation in a
confining medium leads to the suppression patterns summarized in
Figs.\ \ref{psi} and \ref{upsi}.

\par

For a more accurate determination of the quarkonium suppression
patterns, it would be desirable to carry out direct lattice studies of
the color singlet potential and of its quark mass dependence, which
may become important near the critical temperature. Furthermore, to make
contact with nuclear collision experiments, a more precise determination
of the energy density via lattice simulations is clearly needed, as is
a clarification of the role of a finite baryochemical potential.
For the latter problem, lattice studies are so far very difficult;
nevertheless, a recent new approach \cite{fodor} could make such
studies feasible. Finally we note that in hot media the interaction of
a quarkonium state with partonic constituents, in particular gluon
scattering, can obviously also lead to its dissociation
\cite{KS3,Wang}. A study of sequential suppression in such a framework
would certainly be of considerable interest.

\bigskip

\noindent{\bf Acknowledgements}

\bigskip

It is a pleasure to thank F. Karsch and E. Laermann for numerous
helpful discussions. The financial support from DFG under grant
Ka 1198/4-1 and from BMFB under grant 06 BI 902 is gratefully
acknowledged.

\bigskip

\noindent{\bf Appendix}

\bigskip

The NRQCD factorization formula for the inclusive $\u(nS)$ cross section
states \cite{Braaten}:
\begin{eqnarray}
 \sigma[\Upsilon(nS)]_{\rm inc} &=&
 \sigma[b \bar b_1(^3S_1)]
        \langle O_1(^3S_1) \rangle^{\Upsilon(nS)}_{\rm inc}
+ \sum_J  \sigma[b \bar b_1(^3P_J)]
        {\langle O_1(^3P_J) \rangle^{\Upsilon(nS)}_{\rm inc}\over m_b^2}
\nonumber
\\
&& +  \sigma[b \bar b_8(^3S_1)]
        \langle O_8(^3S_1) \rangle^{\Upsilon(nS)}_{\rm inc}
+  \sigma[b \bar b_8(^1S_0)]
        \langle O_8(^1S_0) \rangle^{\Upsilon(nS)}_{\rm inc}
\nonumber
\\
&&
+ \left( \sum_J (2J+1)  \sigma[b \bar b_8(^3P_J)] \right)
        {\langle O_8(^3P_0) \rangle^{\Upsilon(nS)}_{\rm inc}\over m_b^2},
\label{sig-total}
\end{eqnarray}
where $\sigma [b \bar b_{1,8}(^{2 s+1}L_J)]$ are the sort distance
cross sections, $\langle O_{1,8}(^{2s+1}L_J)
\rangle^{\Upsilon(nS)}_{\rm inc}$ are the inclusive NRQCD matrix
elements, which are non-perturbative \cite{Braaten}, and
$m_b$ is the mass of the b-quark. The indices 1 and 8 refers to color
singlet and color octet states, respectively.
In the present analysis the short distance cross sections are not
calculated in perturbation theory, but extracted from the experimental
data on inclusive cross sections ${\sigma(\Upsilon(nS))}_{inc}$
\cite{CDF99}, using the information on matrix elements presented in
\cite{Braaten,Domenech}. The color octet matrix elements for $J=0$
states, i.e. $\langle O_8(^1S_0) \rangle^{\Upsilon(nS)}_{\rm inc}$
$\langle O_8(^3P_0) \rangle^{\Upsilon(nS)}_{\rm inc}$,
are compatible with zero within errors \cite{Braaten}. In fact it was shown
that a good description of the experimental data on bottomonium production
can be obtained simply setting these matrix elements to zero \cite{Domenech}.
Furthermore $\langle O_1(^3P_0) \rangle^{\Upsilon(nS)}_{\rm inc}$ is very small and we
set it to zero in what follows. We also assume that $\sigma[b \bar b_1(^3 P_2)]/\sigma[b \bar
b_1(^3 P_1)]=8.3$, as predicted by perturbative calculations
\cite{Baier,Gastmans}.
With these assumptions, the inclusive cross section for S-states can be
written as
\begin{eqnarray}
 \sigma[\Upsilon(nS)]_{\rm inc} &=&
 \sigma[b \bar b_1(^3S_1)]
        \langle O_1(^3S_1) \rangle^{\Upsilon(nS)}_{\rm inc}+
\sigma[b \bar b_8(^3S_1)]
        \langle O_8(^3S_1) \rangle^{\Upsilon(nS)}_{\rm inc}+\nonumber\\
&&
{\sigma_P\over m_b^2} (8.3 \langle O_1(^3P_2) \rangle^{\Upsilon(nS)}_{\rm inc}+
\langle O_1(^3P_1) \rangle^{\Upsilon(nS)}_{\rm inc}),
\end{eqnarray}
with $\sigma_P=\sigma[b \bar b_1(^3 P_1)]$.
The inclusive color singlet matrix elements are defined as
\begin{eqnarray}
&&
\langle O_1(^3S_1) \rangle^{\Upsilon(nS)}_{\rm inc}=
\sum_{m \ge n} \langle O_1(^3S_1) \rangle^{\Upsilon(nS)} B(\Upsilon(mS) \rightarrow \Upsilon(nS))
\nonumber\\
&&
\langle O_1(^3P_J) \rangle^{\Upsilon(nS)}_{\rm inc}=
\sum_{m \ge n} \langle O_1(^3P_J) \rangle^{\chi_{bJ}(mP)} B(\chi_{bJ}(mP) \rightarrow \Upsilon(nS)),
\end{eqnarray}
where $B(H \rightarrow \Upsilon(nS))$ are the inclusive branching fractions
\cite{Braaten}, and $\langle O_1(^3S_1) \rangle^{\Upsilon(nS)}$ and
$\langle O_1(^3P_J) \rangle^{\chi_{bJ}(mP)}$ are the direct color
singlet matrix elements. The latter can be related to the $b \bar b$
wave function (or its derivative) at the origin \cite{Bodwin}; we
have taken this from \cite{Eichten}, calculated for Buchm\"uller-Tye
potential.
The inclusive color octet matrix elements $\langle O_8(^3S_1) \rangle^{\Upsilon(nS)}_{\rm inc}$
were taken from \cite{Domenech}.
Given the inclusive matrix and the Tevatron data on 1S, 2S and 3S
inclusive production cross sections, one can extract the short distance
cross sections
$\sigma[b \bar b_1(^3S_1)]$, $\sigma[b \bar b_8(^3S_1)]$ and $\sigma_P$.
From the short distance cross sections we can estimate the direct cross
section for different quarkonium states,
\begin{eqnarray}
&&
\sigma_d[\Upsilon(nS)]=\sigma[b \bar b_1(^3S_1)]
        \langle O_1(^3S_1) \rangle^{\Upsilon(nS)}+
\sigma[b \bar b_8(^3S_1)]
        \langle O_8(^3S_1) \rangle^{\Upsilon(nS)}+\nonumber\\
&&
\sigma_d[\chi_{bJ}(nP)]=\sigma[b \bar b_1(^3 P_J)] \langle O_1(^3P_J)
\rangle^{\chi_{bJ}(nP)}/m_b^2.
\label{sigma-direct}
\end{eqnarray}
In the direct cross section for the $\chi_{bJ}$ states we have neglected
the color octet contribution which is proportional to $\langle
O_8(^3S_1) \rangle^{\chi_{bJ}(nP)}$, in accordance with the results
of \cite{Domenech}, where it was show that $\chi_{bJ}$ states are
dominantly produced by a color singlet mechanism. Also the analysis of
Ref. \cite{Braaten} shows that the matrix elements $\langle O_8(^3S_1)
\rangle^{\chi_{bJ}(nP)} \simeq 0$.
To complete our analysis we have to estimate the direct color octet
matrix elements entering in Eq.\ (\ref{sigma-direct}). With the
assumption that $\langle O_8(^3S_1)\rangle^{\chi_{bJ}(nP)}=0$ we can
write for the direct color octet matrix elements
\begin{eqnarray}
\langle O_8(^3S_1) \rangle^{\Upsilon(1S)}_{\rm inc}&=&\langle O_8(^3S_1) \rangle^{\Upsilon(1S)}+
B(\Upsilon(2S) \rightarrow \Upsilon(1S)) \langle O_8(^3S_1) \rangle^{\Upsilon(2S)}+\nonumber\\
&&
B(\Upsilon(3S) \rightarrow \Upsilon(1S)) \langle O_8(^3S_1) \rangle^{\Upsilon(3S)}+\nonumber\\
\langle O_8(^3S_1) \rangle^{\Upsilon(2S)}_{\rm inc}&=&\langle O_8(^3S_1) \rangle^{\Upsilon(2S)}+
B(\Upsilon(3S) \rightarrow \Upsilon(2S)) \langle O_8(^3S_1) \rangle^{\Upsilon(3S)}+\nonumber\\
\langle O_8(^3S_1) \rangle^{\Upsilon(3S)}_{\rm inc} &\approx& \langle O_8(^3S_1) \rangle^{\Upsilon(3S)}
\end{eqnarray}
Finally, multiplying the direct cross section for different bottomonium
states by the corresponding inclusive branching fractions from
\cite{Braaten}, we obtain the feed-down fractions given in Table 2.

\vspace*{4cm}
\centerline{\bf Figures }

\end{document}